\documentclass[aps,pra,superscriptaddress,showfacs,twocolumn,showpacs]{revtex4}
\usepackage{amsmath}
\usepackage{graphicx}
\usepackage{amssymb}
\usepackage{bbm, color}

\newcommand*{\tn}[1]{{\textnormal{#1}}}
\newcommand{\ket}[1]{\left|#1\right>}
\newcommand{\bra}[1]{\left<#1\right|}
\newcommand{\inn}[2]{\left<#1|#2\right>}

\newcommand*{\1}{{\mathbbm{1}}}

\begin{document}

\title{
Non-Hermiticity and conservation of orthogonal relation in dielectric microcavity}

\author{Kyu-Won Park}
\affiliation{Department of Physics and Astronomy, Seoul National University, Seoul 08826, Korea}
\author{Songky Moon}
\affiliation{Department of Physics and Astronomy, Seoul National University, Seoul 08826, Korea}
\author{Hyunseok Jeong}
\affiliation{Center for Macroscopic Quantum Control, Department of Physics and Astronomy, Seoul National University, Seoul 08826, Korea}
\author{Jaewan Kim}
\affiliation{School of Computational Sciences, Korea Institute for Advanced Study, Seoul 02455, Korea}
\author{Kabgyun Jeong}
\email{kgjeong6@snu.ac.kr}
\affiliation{IMDARC, Department of Mathematical Sciences, Seoul National University, Seoul 08826, Korea}
\affiliation{Center for Macroscopic Quantum Control, Department of Physics and Astronomy, Seoul National University, Seoul 08826, Korea}
\affiliation{School of Computational Sciences, Korea Institute for Advanced Study, Seoul 02455, Korea}

\date{\today}

\begin{abstract}
Non-Hermitian properties of open quantum systems and their applications have attracted much attention in recent years. While most of the studies focus on the characteristic nature of non-Hermitian systems, here we focus on the following issue: A non-Hermitian system can be a subsystem of a Hermitian system as one can clearly see in Feshbach projective operator (FPO) formalism. In this case, the orthogonality of the eigenvectors of the total (Hermitian) system must be sustained, despite the eigenvectors of the subsystem (non-Hermitian) satisfy the bi-orthogonal condition. Therefore, one can predict that there must exist some remarkable processes that relate the non-Hermitian subsystem and the rest part, and ultimately preserve the Hermiticity of the total system. In this paper, we study such processes in open elliptical microcavities. The inner part of the cavity is a non-Hermitian system, and the outer part is the coupled bath in FPO formalism. We investigate the correlation between the inner- and the outer-part behaviors associated with the avoided resonance crossings (ARCs), and analyze the results in terms of a trade-off between the relative difference of self-energies and collective Lamb shifts. These results come from the conservation of the orthogonality in the total Hermitian quantum system.
\end{abstract}
\maketitle

\section{Introduction} \label{intro}
Hermicity of physical observables is one of the basic principles in quantum mechanics. For given Hermitian operator, all of its eigenvalues are real, and its eigenvectors corresponding to different eigenvalues are orthogonal to each other.  On the other hand, a non-Hermitian system, which is related to openness has complex eigenvalues and its eigenvectors satisfy the bi-orthogonal relation. Recently, various non-Hermitian systems and their properties have been extensively studied theoretically as well as experimentally~\cite{M11,F58,R09,CK09,R10,PRSB00,CW15}, especially in the fields of avoided resonance crossings (ARCs)~\cite{RLK09,W06,SGLX14,SGWC13,WH06,RPPS00,BP99,B96}, exceptional points (EPs)~\cite{K66,SKM+16,CK17}, $PT$-symmetric Hamiltonian systems~\cite{BB98,GS09,XC17,ZGW+17}, phase rigidity~\cite{BRS07}, bi-orthogonal relations~\cite{C06,B14,CS03,L09}, and optical chirality around the EP~\cite{W14,W11,ZGL+11}. Since the Hermitian system and the non-Hermitian system has quite different properties to each other, they are often considered being distinct and independent from each other.

For an open system, one can consider a total system composed of the open system and a bath interacting with the system. In other words, the total system, which is Hermitian, is decomposed into two orthogonal subspaces; one is a non-Hermitian (sub-)system, and the other is a bath coupled to it. This decomposition is known as the Feshbach projective operator (FPO) formalism~\cite{F58,R09,M11,D00}. It was first introduced by Feshbach in 1958 for describing a decay process in nuclear physics, and has been extended to other systems such as quantum dots~\cite{R09}, microwave cavities~\cite{PRSB00}, and dielectric microcavities~\cite{PKJ16,PKJ16+}. In studies employing FPO formalism, a focus is usually made on the non-Hermitian part rather than the coupled bath. However, it should be noted that the Hermitian properties of the total system is maintained despite of the existence of a non-Hermitian (sub-)system. There must occur some correlations in the system-bath interaction that preserve the Hermiticity of the total system.

In this paper, we study such correlations for two-dimensional dielectric microcavities. Even though the dielectric microcavities are classical systems, they can be a good platform to study the wave-mechanical properties of quantum mechanics due to an isomorphic nature of wave equations between optics and quantum mechanics~\cite{RBM12+,DMB16+}. The total system consists of the inner-part of the cavity (non-Hermitian) and the outer-part of the cavity (bath). We investigate the correlations between the inner- and the outer-parts of the wavefunctions in the context of the avoided resonance crossing (ARC) as a function of the eccentricity of ellipse, in which the resonances of the non-Hermitian part are strongly interacting and thus undergoing dramatic changes. The ARC in an open system is a natural extension of the avoided level crossing (ALC)~\cite{LL65} occuring in a closed system. Moreover, we adopt ellipses as the boundary shape of the microcavities in order to see the manifestation of the openness nature clearly; For a closed elliptic billiard, since there are no internal interactions causing the real-valued energy repulsion~\cite{H10,T89}, no ALC occurs (the elliptic billiard may have the long-range avoided crossings known as Demkov-type in some cases~\cite{KK17}, but we only consider Landau-Zener-type avoided crossing here). On the other hand, in an open elliptic system, there can occur ARCs~\cite{W06}. 

Next, we employ the Lamb shift to understand the findings in the above investigations. The Lamb shift formally describes the openness nature of a quantum system in the system-bath coupling~\cite{LR47,WKG04,N10}. Originally, this concept was known as a small difference in energy levels of a hydrogen atom in quantum electrodynamics, caused by the vacuum fluctuations~\cite{LR47}. However, it is recently found that there are two types of the Lamb shift. One is, so-called, the self-energy and the other is the collective Lamb shift~\cite{SS10,RWSS12,R13}. The self-energy is simply known as the Lamb shift in atomic physics. It is an energy-level shift arising from individual interaction of energy-level with its bath. On the other hand, the collective Lamb shift is an energy-level shift due to the interaction of energy-levels with each other via the bath.  The system-bath coupling in non-Hermitian Hamiltonian is analogy to Lamb shift. However, in the microcavity society, we conventionally call it as Lamb shift. The diagonal (off-diagonal) components of non-Hermitian Hamiltonian correspond to the self-energy (collective Lamb shift)~\cite{R13,PKJ16,PKJ16+}. Our former works considered only self-energy in circular and elliptic dielectric microcavity~\cite{PKJ16,PKJ16+}. However, in this study, we consider both collective Lamb shift and self-energy, and show that an interplay between collective Lamb shift and relative difference of self-energies determines the essential features of ARC. Moreover, it will be shown that this interplay comes from the orthogonality of wavefunctions for the total Hermitian Hamiltonian.

\section{Correlation of the system wavefunction with that of the bath} \label{CvsAC}
Let $\mathcal{H}_T$ be a total Hermitian Hamiltonian with real eigenvalues $\mu_T$, and $\ket{\mu_T}_{SB}$ represents the corresponding eigenvector of $\mu_T$ for the total system $SB$.
Any wavefunctions of a Hermitian Hamiltonian corresponding to the total Hilbert space can be decomposed into two orthogonal subspaces, a close quantum system $S$ and a bath part $B$, by using the Feshbach projective operator $p_{S}$, $p_{B}$: $p_{S}$ is a projection operator onto the closed quantum system $S$ whereas $p_{B}$ is a projection onto the bath $B$ with $p_{S}+p_{B}=\1_{T}$ and $p_{S}p_{B}=p_{B}p_{S}=0$. The operator $\1_{T}$ is an identity operator for the total (system-bath) space. By using these projection operators, we can define useful matrices such as $H_{S} = p_{S}\mathcal{H}_{T}p_{S},~H_{B} = p_{B}\mathcal{H}_{T}p_{B},~\mathcal{V}:= V_{SB} = p_{S}{\mathcal{H}}_{T}p_{B}$, and $\mathcal{V}^{\dag} := V_{BS} = p_{B}{\mathcal{H}}_{T}p_{S}$.  Then, the total Hamiltonian is given by a matrix $\mathcal{H}_T=H_S+H_B+\mathcal{V}+\mathcal{V}^\dagger$ and the total eigenvector on $SB$ is also given by $\mu_{SB}=\mu_{S}+\mu_{B}.$  After rearranging those equations, We obtain the following non-Hermitian Hamiltonian
 \begin{align}
 \hat{H}_\tn{eff}=H_S+\mathcal{V}\mathcal{G}_B^{\triangleright}\mathcal{V}^\dagger
 \end{align}
with $j$-th complex energy eigenvalue $\nu_{j}$ and its eigenvector $\ket{\psi_{j}}$, and $\mathcal{G}_B^{\triangleright}$ denotes the out-going Green's function defined by $\mathcal{G}_B^{\triangleright}:=(\beta^+-H_B)^{-1}$. Notice that $\hat{H}_\tn{eff}\ket{\psi_{j}}_\tn{eff}=\nu_{j}\ket{\psi_{j}}_\tn{eff}$ and $\beta^+$ is an eigenvalue of $H_B$ with small positive imaginary part ($\eta$) added for out-going state, i.e., $\beta^+=\beta+i\eta$ in the limit of $\eta\to0^+$. Then, for each $j$, we have
\begin{align}
\ket{\mu_{T,j}}_{SB}=\ket{\psi_{j}}_S+\ket{\chi_{j}}_B,
\end{align}
where $\ket{\psi_{j}}_S$ is an eigenvector of non-Hermitian Hamiltonian $\hat{H}_\tn{eff}$ in open quantum system and $\ket{\chi_j}_{B}$ is a resonance tail only localized in bath omitting the homogeneous solution (or the plane-wave term) $\ket{\beta}_B$ (See details in Appendix A). In the case of dielectric microcavity, $\ket{\psi_{j}}_S$ correspond to eigenmode of the inner part of the cavity and $\ket{\chi_{j}}_B$ to the emission pattern of the outer-part of it~\cite{PKJ16}, respectively.
Then the inner product between two eigenvectors of the total Hamiltonian will be given by
\begin{align}
{\mathstrut}_{S}\!\inn{\psi_{j}}{\psi_{k}}_S+{\mathstrut}_{B}\!\inn{\chi_{j}}{\chi_{k}}_B=\delta_{jk}, \label{eq:Soverlapp}
\end{align}
which directly follows from the fact that $\forall j,k$, ${{}_{}}_{SB}\!\inn{\mu_{T,j}}{\mu_{T,k}}_{SB}=\delta_{jk}$, i.e., it must be always orthogonal. Thus, ${\mathstrut}_{S}\!\inn{\psi_{j}}{\chi_{k}}_B=0$ and ${\mathstrut}_{B}\!\inn{\chi_{j}}{\psi_{k}}_S=0$: It comes from that the subspaces $S$ and $B$ are mutually orthogonal. Note that the LHS of Eq.~(\ref{eq:Soverlapp}) is non-trivial, since the eigenvectors of the non-Hermitian Hamiltonian are bi-orthogonal, i.e., $\langle{{\psi}^{L}_{j}}|{{\psi}^{R}_{k}}\rangle= \langle{{\psi}^{*}_{j}}|{{\psi}_{k}}\rangle=\delta_{jk}$. (See also details in Appendix B.) The Eq.~(\ref{eq:Soverlapp}) means that when the inner product of eigenvectors of $\hat{H}_{\tn{eff}}$, ${\mathstrut}_{S}\!\inn{\psi_{j}}{\psi_{k}}_{S}$, increases, that of resonance tail, ${\mathstrut}_{B}\!\inn{\chi_{j}}{\chi_{k}}_{B}$, must vary in a way to cancel out ${\mathstrut}_{S}\!\inn{\psi_{j}}{\psi_{k}}_{S}$ for preserving orthogonal relation in total Hilbert space. In this way, the inner-part of the microcavity and the outer-part of it have an intimate correlation.

\subsection{Inner-part behaviors of microcavity}
\begin{figure*}
\centering
\includegraphics[width=18.5cm]{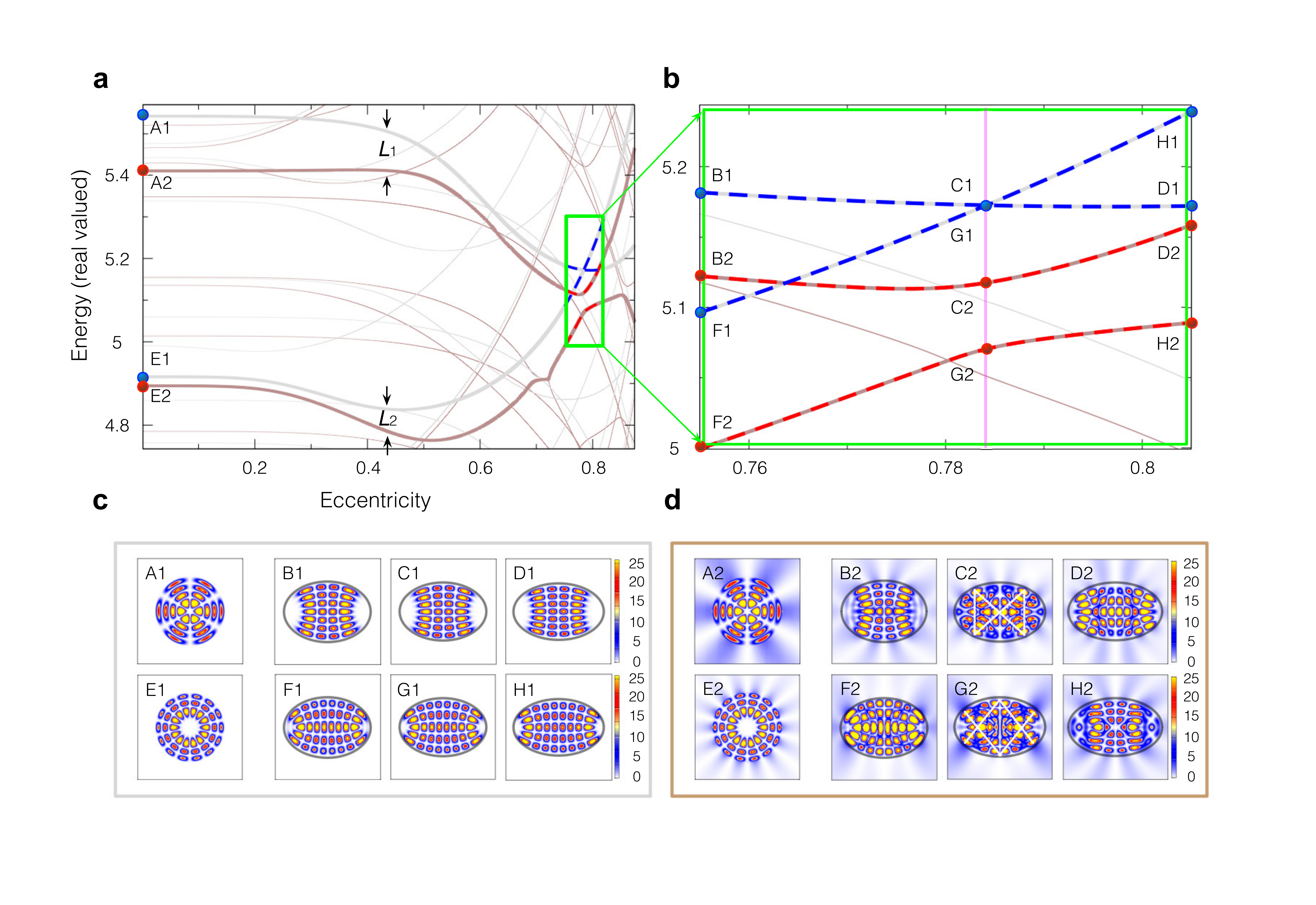}
\caption{{\bf The eigenvalue trajectories in terms of real-valued energies and their intensity-plots of the wavefunctions in the elliptic system.} (\textbf{a}) The eigenvalue trajectories in terms of real-valued energies for the elliptic billiards and elliptic dielectric microcavities are represented by grey solid lines and brown solid lines as $e$ is varied. The thick grey and brown solid lines of A1 and A2 and those of E1 and E2 correspond to radial quantum number and angular quantum number of ($\ell=4, m=3$) and ($\ell=3, m=7$), respectively.
(\textbf{b}) Extension of the green solid box in Fig.~\ref{Figure-1}\textbf{a}, where blue dashed lines for the elliptic billiard are crossing near at the eccentricity $e\approx 0.782$ and red dashed lines for the microcavity occur avoided crossing near at the similar point (i.e., $e\approx 0.782$).  (\textbf{c}) Intensity plots of wavefunctions for the blue dots (from A1 to H1). They are entirely localized on the interior of the elliptic boundary. (\textbf{d}) Intensity plots of wavefunctions for the red dots (A2 to H2). They exit on the total space. The figures of C2 and G2 are symmetric and antisymmetric coherent superposition of B2 and F2, respectively.}
\label{Figure-1}
\end{figure*}

In order to study the correlations above, we have obtained the eigenvalues and their intensity-plots of the eigenfunctions in elliptic cavity by using the boundary element method (BEM) below~\cite{W03}. There are two kinds of the real-valued energies in the eigenvalue trajectories represented by grey solid and brown solid curves in Fig.~\ref{Figure-1}\textbf{a}. The grey solid curves are the real-valued energies of eigenvalue trajectories for the elliptic billiard belonging to integrable system, whereas the brown solid curves are those for the dielectric microcavity or open quantum system.

The upper thick solid grey curve (A1) shows an eigenvalue trajectory with radial quantum number of $\ell=3$ and angular quantum number of $m=7$ varying from circular to elliptic  billiard depending on the eccentricity $e$, which is defined by $e=\sqrt{1-(b/a)^2}$, where $a$ and $b$ are the major and minor axes of the ellipse, respectively. The upper thick solid brown curve (A2) shows that of an elliptic dielectric microcavity with refractive index $n=3.3$ for InGaAsP. On the other hand, the lower thick solid grey and brown curves also show the eigenvalue trajectories with different quantum number $\ell=4$ and $m=3$. They start from the eccentricity $e=0$ (circle) to $e\approx0.805$ (ellipse) showing real-valued energies crossing for the billiard but avoided crossing for microcavity around $e\approx0.782$. Notice those facts in Fig.~\ref{Figure-1}\textbf{b}, i.e., the enlarged version of a green solid box in Fig.~\ref{Figure-1}\textbf{a}.

\begin{figure}
\centering
\includegraphics[width=8.5cm]{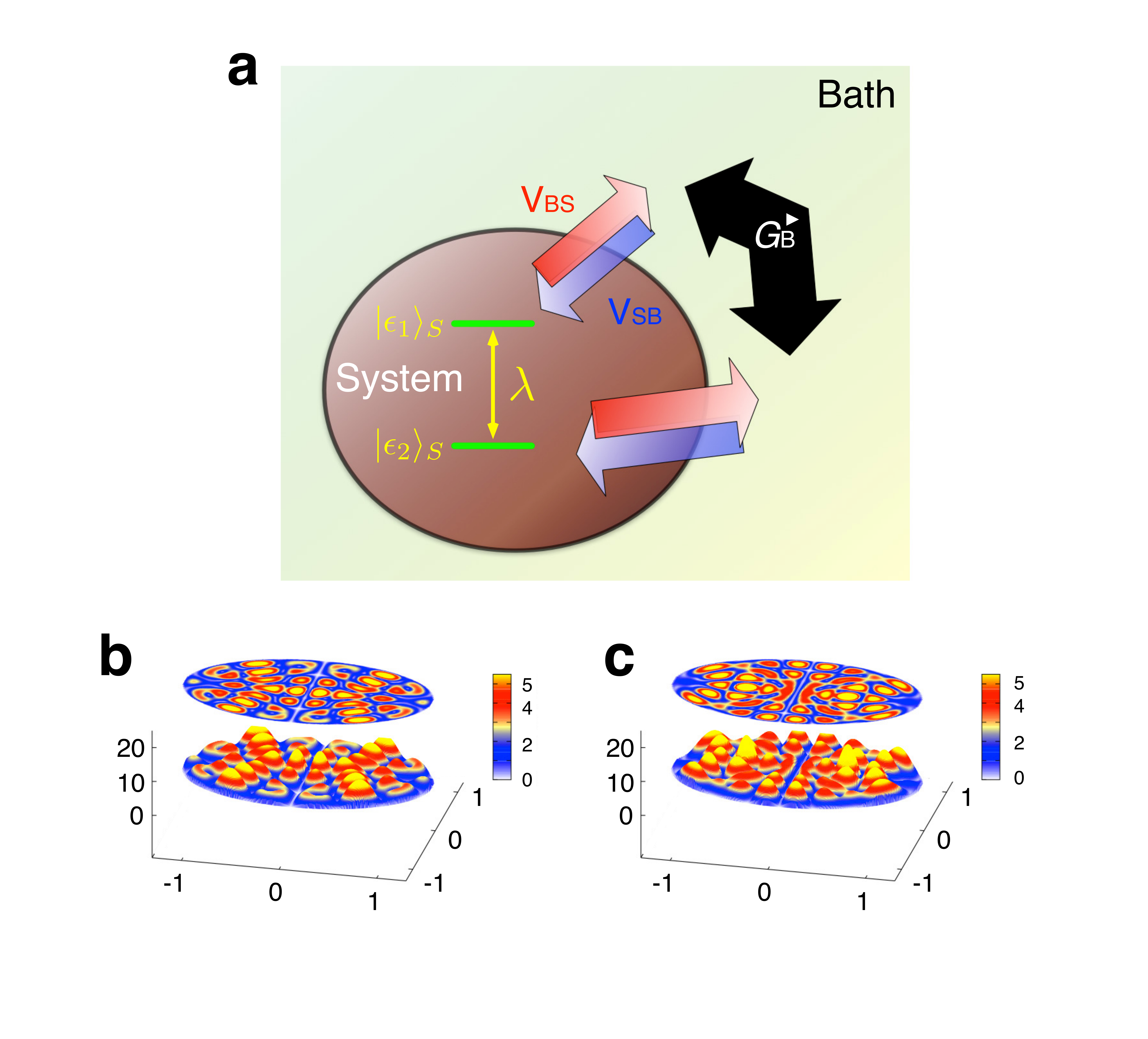}
\caption {{\bf The schematic diagram for the non-Hermitian Hamiltonian and its superposed wavefunctions near at the ARC}. (\textbf{a}) There are three kinds of Hamiltonian toy models controlled by system-bath interactions $V_{BS}$ or $V_{SB}$ and out-going Green's function in bath $\mathcal{G}_B^{\triangleright}$ (as in Appendix A). Here, three kinds of toy models are presented. First, $\hat{H}_\tn{eff}=H_S$: There are not any system-bath interaction. In this case, the non-Hermitian Hamiltonian is just equal to Hermitian Hamiltonian. Second, $\hat{H}_\tn{eff}=(\hat{H}_\tn{eff})_D$: The state of closed quantum system $\ket{\epsilon_j}_S$ can be leaky out to bath by $V_{BS}$ and then $\mathcal{G}_B^{\triangleright}$ guides to $\ket{\epsilon_j}_S$ again, and penetrate into $\ket{\epsilon_j}_S$ by $V_{SB}$ resulting in the self-energy $\gamma_{jj}$. Third, $\hat{H}_\tn{eff}=(\hat{H}_\tn{eff})_D+(\hat{H}_\tn{eff})_V$: The Green's function $\mathcal{G}_B^{\triangleright}$ can guide to not only same state $\ket{\epsilon_j}_S$ but also to different  state ($\ket{\epsilon_j}_S\neq\ket{\epsilon_k}_S$) resulting in the collective Lamb shift $\gamma_{jk}$ where $j\neq k$.
(\textbf{b}) The intensities of the symmetric superposed state through $\mathcal{G}_B^{\triangleright}$ with wavefunctions for F2 and B2 in Fig.~\ref{Figure-1}\textbf{d} is obtained. (\textbf{c}) The intensities of anti-symmetric superposed states through $\mathcal{G}_B^{\triangleright}$ with wavefunctions for F2 and B2 in Fig.~\ref{Figure-1}\textbf{d} is obtained. The Fig.~\ref{Figure-2}\textbf{b} and Fig.~\ref{Figure-2}\textbf{c} denote the bowtie- and $V(\Lambda)$-type modes in the microcavity, respectively. See details the differences in Subsec.~\ref{Conservation law}.
} 
\label{Figure-2}
\end{figure}

There are intensity plots of wavefunctions corresponding to the eigenvectors of elliptic billiards, in Fig.~\ref{Figure-1}\textbf{c} for blue dots (from A1 to H1). All of them are entirely localized on the interior regions of the elliptic boundary, and have well-defined quantum numbers by an elliptic and hyperbolic coordinate. Fig.~\ref{Figure-1}\textbf{d} presents that intensity plots of wavefunctions for red dots (from A2 to H2). They exist on the total space not inside elliptic boundary and do not have well-defined quantum number around ARC region. However, in order to focus on the properties of non-Hermitian Hamiltonian, we only deal with inner parts of microcavity in this section. The interior wavefunctions of blue and of red dots are quite similar to each other away from ARC region (self-engergy), while those of blue and of red dots are different around ARC region (collective Lamb shift). The wavefunctions B1, C1, D1, F1, G1, and H1 retain their own shapes as stable bouncing-ball-type modes whereas the wavefunctions for C2 and G2 show pronounced deformation from bouncing-ball-type modes and [B2 $\Leftrightarrow$ H2, F2 $\Leftrightarrow$ D2] undergo mode exchanges due to coherent superpositions caused by the strong-interaction. The wavefunctions C2 and G2 are symmetric and anti-symmetric superpositions with B2 and F2 and  they are known as bowtie- and $V(\Lambda)$-type modes. These kinds of behaviors can be explained by three different types of Hamiltonians in Fig.~\ref{Figure-2}.

\subsection{Outer-part behaviors of microcavity}

\begin{figure*}
\centering
\includegraphics[width=18cm]{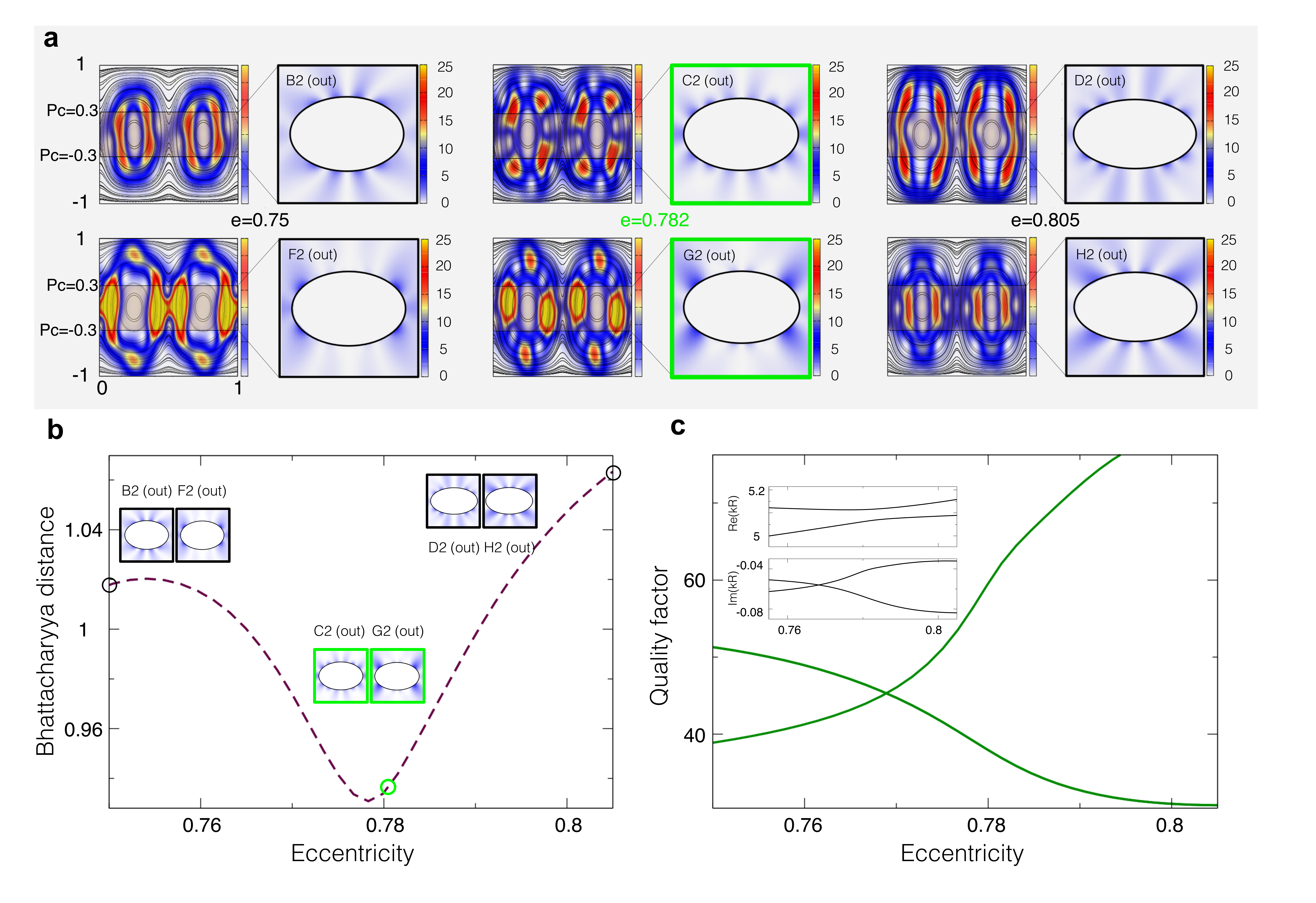}
\caption {{\bf Husimi distributions of decay channels and the Bhattacharyya distance}. (\textbf{a}) Husimi probability distributions and resonance tails of resonance modes in Fig.~\ref{Figure-1}\textbf{d} at eccentricity $e=0.75$, $e=0.782$, and $e=0.805$, respectively. The shaded regions of each Husimi probability distributions determine the resonance tails (i.e., decay channels). (\textbf{b}) The Bhattacharyya distance, $d_B$, between two decay channels as a function of the eccentricity $e$. The extremal point of $d_B$ is near $e\approx0.78$ where the bouncing-ball-type modes are mixed at bowtie- and $V(\Lambda)$-type modes, respectively. We attribute a small deviation between the extremal point of the Bhattacharyya distance ($d_B$) and the center of ARC ($e_C$) to asymmetric two red dashed curves for the ARC in Fig.~\ref{Figure-1}\textbf{b}. (\textbf{c}) Two quality factors, $Q$'s, for red dashed curves in Fig.~\ref{Figure-1}\textbf{b}. They show same order of magnitude. In the inset \textbf{c}, there are real ($\tn{Re}(kR)$) and imaginary values ($\tn{Im}(kR)$) of the complex energy in eigenvalue trajectories as the eccentricity $e$ is varied. The quality factors are obtained from these two values.}
\label{Figure-3}
\end{figure*}

We plot Husimi probability distributions and the emission pattern of the outer-part of each resonance modes in Fig.~\ref{Figure-3}\textbf{a} and they correspond to the resonance modes in Fig.~\ref{Figure-1}\textbf{d} with the eccentricity $e=0.75$, $e=0.782$, and $e=0.805$. It is well-known that the Husimi distributions below critical lines ($P_c$'s) determine the shape of the resonance tails (i.e., the emission pattern)~\cite{LYMLA07,SLY+07,CCSN00,SSWS+10}. The Fig.~\ref{Figure-3}\textbf{a} well describes this fact, i.e., the shaded grey regions in each figure between critical lines ($P_c=[-0.3,0.3]$) and the resonance tails provide our claims. Therefore, we can compare the Husimi distributions below critical lines instead of resonance tails themselves which extend up to infinity.

More precisely, we can depict a Bhattacharyya distance $d_B$~\cite{B43,MMPZ08} between two decay channels (or resonance tails) as a function of the eccentricity $e$ in Fig.~\ref{Figure-3}\textbf{b}. The $d_B$ in general measures a similarity between two probability distributions $p(x)$ and $q(x)$. Any given probability distributions, it is defined as
\begin{align}
d_B(p(x),q(x))=-\tn{ln}[K_{B}(p(x),q(x))],
\end{align}
where the factor $K_{B}$ (Bhattacharyya coefficient) is given by $K_{B}(p(x),q(x))=\int\sqrt{p(x)q(x)}dx.$ Together with $d_B$, we measure the degree of sharing of common decay channels by comparing the Husimi probability distributions below the critical lines. The extremal point of $d_B$ lies at $e\approx0.78$ where the bouncing-ball-type modes (black circles) are mixed at bowtie-type and $V(\Lambda)$-type modes. It shows that the two mixed modes (i.e., bowtie- and $V(\Lambda)$-type mode; green circle in Fig.~\ref{Figure-3}\textbf{b}) resemble each other more than the two bouncing ball-type modes, and their imaginary part of the complex eigenvalues are crossing simultaneously. The two different quality factors $Q$ are defined by $\mu_j/2\omega_j$ (from the equation~(\ref{eq:evHeff}) in the Appendix); $Q$'s are the values obtained from the red dashed curves in Fig.~\ref{Figure-1}\textbf{b} and they are shown in Fig.~\ref{Figure-3}\textbf{c}. These quality factors are obtained from inset \textbf{c} and they have same order of magnitude.

\subsection{Correlation of the conservation with Lamb shift and avoided crossing} \label{Conservation law}

\begin{figure*}
\centering
\includegraphics[width=17cm]{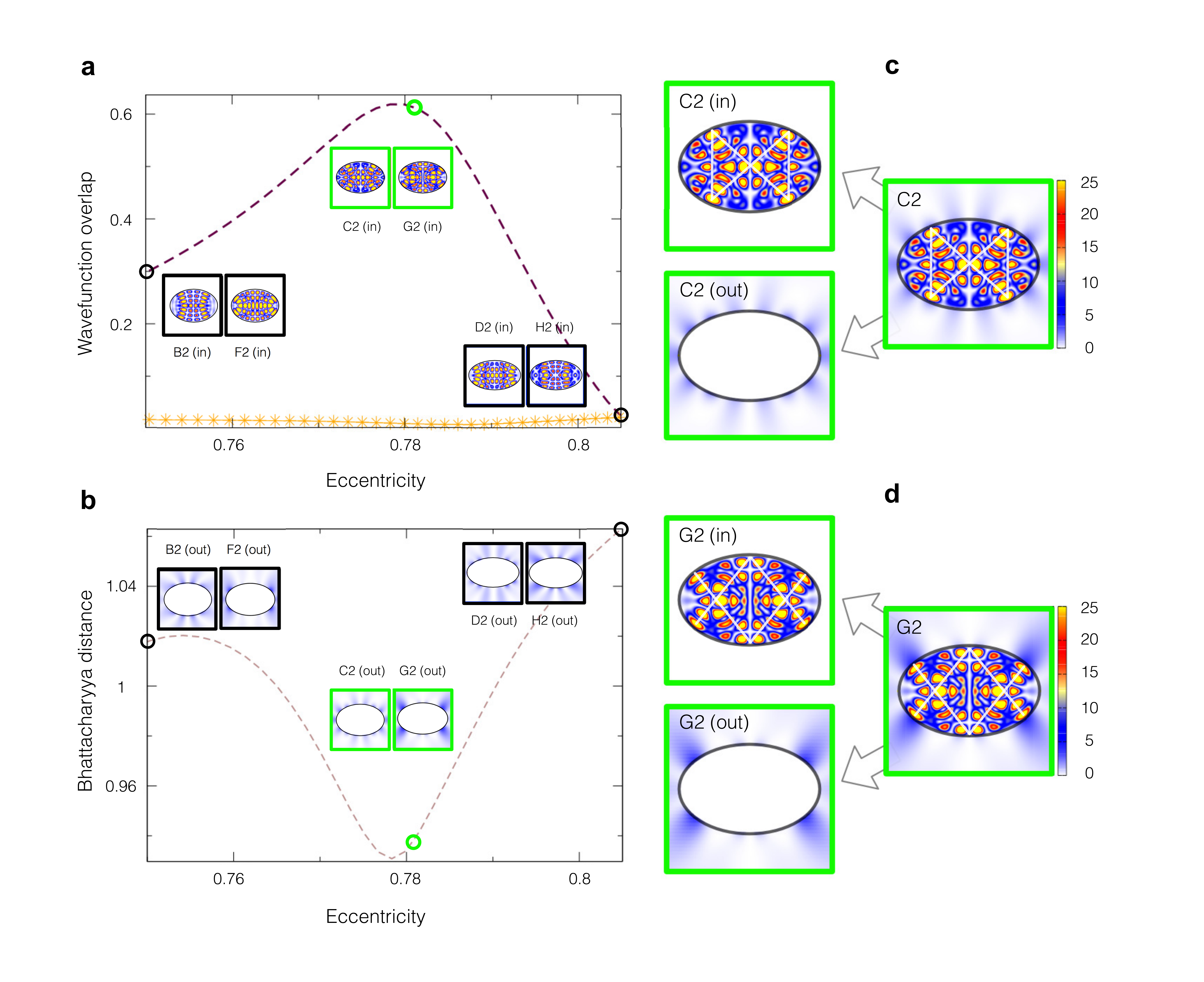}
\caption {{\bf There exists an explicit trade-off between the wavefunction overlap and the Bhattacharyya distance}. (\textbf{a}) An overlap for inner parts of resonance modes in Fig.~\ref{Figure-1}\textbf{d}. At the center of avoided resonance crossing, i.e., where the two bouncing-ball-type modes are mixed at bowtie- and $V(\Lambda)$-type modes, the wavefunction-overlap has maximal value at the center of ARC, and decrease as deviating from the center of ARC. A orange star line in bottom clearly shows a bi-orthogonal relation: That is, $\langle{{\psi}^{L}_{j}}|{{\psi}^{R}_{k}}\rangle=\langle{{\psi}^*_{j}}|{{\psi}_{k}}\rangle=\delta_{jk}$. (\textbf{b}) The Bhattacharyya distance $d_B$ of Fig.~\ref{Figure-3} is shown again for comparing \textbf{a}. We can easily check that the wavefunction overlap for the inner-part of resonance modes and the Bhattacharyya distance for the outer-part of them have almost symmetric shape to each other. (\textbf{c}, \textbf{d}) It precisely represent that the wavefucntions defined in total space can be divided into two parts: (i) inner parts of wavefunctions as the eigenfucntions of non-Hermitian Hamiltonian. (ii) outer parts of wavefunctions as the resonance tails.}
\label{Figure-4}
\end{figure*}

The correlation in conservation of the orthogonal relation for dielectric microcavity is shown in Fig.~\ref{Figure-4} as a trade-off between the wavefunctions' overlap for inner-part of the microcavity and the Bhattacharyya distance for outer-part of the cavity. The overlap in Fig.~\ref{Figure-4}\textbf{a} is obtained from $\sqrt{|\inn{\psi_j}{\psi_k}|}$ as a function of the eccentricity $e$. It shows a parabolic curve whose maximum value is near at the center of the avoided resonance crossing ($e=e_C$) where two bouncing-ball-type modes are mixed at bowtie- and $V(\Lambda)$-type modes, respectively and decrease as deviating from the $(e_{C})$ returning to bouncing-ball-type modes. An orange star line in the bottom in Fig.~\ref{Figure-4}\textbf{a} clearly shows a bi-orthogonal relation:$\langle{{\psi}^{L}_{j}}|{{\psi}^{R}_{k}}\rangle=\langle{{\psi}^*_{j}}|{{\psi}_{k}}\rangle=\delta_{jk}$. Instead of the inner product between resonance tails at infinity ${\mathstrut}_{B}\!\inn{\chi_{j}}{\chi_{k}}_B$, we use the Bhattacharyya distance $d_B$ in Fig.~\ref{Figure-4}\textbf{b} to quantify the similarity between two resonance tails and it shows a symmetric curve with respect to the overlap curve in Fig.~\ref{Figure-4}\textbf{a}. This symmetric shape between the overlap curve and the Bhattacharyya distance shows certain correlation between inner-part of and outer-part of the microcavity as a result of the conservation of the orthogonal relation for the total Hilbert space. In this paper, we insist that this correlation can be used to understand the structure of the Lamb shift and avoided crossing relating to the non-Hermitian Hamiltonian itself.

As mentioned in Appendix A for the details of non-Hermitian (effective) Hamiltonian, let us now write $\hat{H}_{\tn{eff}}:=H_S+\Gamma$ as matrix form with respect to the eigenbasis of $H_{S}$ explicitly to understand the Lamb shift and avoided crossing in Fig.~\ref{Figure-1}, under the continuous change parameter $e$, as
\begin{align}
\hat{H}_{\tn{eff}}(e)
&=\begin{pmatrix}
\epsilon_{1}(e) & \lambda_{1}(e) \\
\lambda_{2}(e) & \epsilon_{2}(e)
\end{pmatrix} +
\begin{pmatrix}
\gamma_{11}(e) & \gamma_{12}(e) \\
\gamma_{21}(e) & \gamma_{22}(e)
\end{pmatrix}.
\end{align}
with $\lambda:=\lambda_1=\lambda_2^*$ (Note that $H_S$ is Hermitian). The first matrix represents Hamiltonian of a closed quantum billiard system and second one does system-bath interaction. Since $H_{S}$ is a Hamiltonian of a closed quantum billiard system, it is a Hermitian matrix and has real eigenvalues (See Fig.~\ref{Figure-2}). Especially, if the billiard systems are one of the integrable systems~\cite{H10,T89}, there are no interactions in it so that the off-diagonal term (internal coupling) $\lambda_j(e)$ must be vanished while giving $\epsilon_j(e)=\epsilon_j$. For convenience, we omit the parameter $e$ when the meaning of the variable $e$ is obvious. When the eccentricity $e$ is equal to a crossing point, i.e., $e=e_X$, the eigenvalues of $H_{S}$ are degenerated with $\epsilon_1(e_{X})=\epsilon_2(e_{X})$, but not eigenvectors. This explains the process of thick grey curves in Fig.~\ref{Figure-1}. On the other hand, the second matrix is quite different from that of the first one. It is a Hamiltonian due to system-bath coupling with complex entries. When the Hamiltonian represents time reversal system, it becomes symmetric so that $\gamma_{12}$ is same to $\gamma_{21}$~\cite{L09}. Therefore, the final form of matrix in our case under the parameter $e$ is given by
\begin{align}
\hat{H}_{\tn{eff}}(e)
=\begin{pmatrix}
\epsilon_1+\gamma_{11} & \Re(\gamma')+i\Im(\gamma') \\
\Re(\gamma')+i\Im(\gamma') & \epsilon_2+\gamma_{22} \label{eq:symHeff}
\end{pmatrix},
\end{align}
where $\gamma':=\gamma_{12}=\gamma_{21}$. Then, there are primarily three types of interactions depending on $\gamma'$~\cite{BP99,RPPS00,B96}. First $\gamma$'s are pure real leading to the repulsion of real parts and the crossing of imaginary parts in the complex energy. Second, $\gamma'$s are pure imaginary leading to a repulsion of imaginary parts and a crossing of real parts. Third, $\gamma'$s are the complex number resulting in the repulsion of both parts simultaneously. Since we here deal with only the first case (strong coupling), we must set $\gamma'$ to be $\Re(\gamma')$. We can straightforwardly diagonalize the Eq.~(\ref{eq:symHeff}) above, and then we obtain a following equation satisfying
\begin{align}
\nu_\pm=&\frac{\nu_1^D+\nu_2^D}{2}\pm z_\nu, \label{eq:evHeff2}
\end{align}
where $z_\nu=\big[[\epsilon_1- \epsilon_2+\Re(\gamma_{11})-\Re(\gamma_{22})+i\{\Im(\gamma_{11})-\Im(\gamma_{22})\}]^2+4\Re^2(\gamma')\big]^{1/2}$.
Note that the the diagonal eigenvalues $\nu_{1,2}^D:=\epsilon_{1,2} +\Re(\gamma_{11,22})+i\Im(\gamma_{11,22})$ denote the eigenvalues of the system's Hamiltonian $(\hat{H}_\tn{eff})_D$. Here, we assume that $\Im(\gamma_{11})-\Im(\gamma_{22})$ is negligible, i.e., $\Im(\gamma_{11})-\Im(\gamma_{22})\approx 0$. This is a reasonable assumption that the real part of the complex energy is repulsion, but the imaginary part is crossing with $|\Im(\gamma_{jj})|\ll1$~\cite{RPPS00,BP99,B96} and it considers $e_{X}\approx e_{C}$. Then, the repulsion of the real part near the center of ARC ($e=e_C$) is given by
\begin{equation} \label{eq:2znu}
2z_\nu\approx 2\sqrt{[\Re(\gamma_{11})-\Re(\gamma_{22})]^2+4\Re^2(\gamma')}.
\end{equation}

The first term of the square root corresponds to the relative difference of self-energy, and the second term does to the collective Lamb shift, respectively. It is important to note that the physical meaning of the relation, $\Re(\gamma_{jj})-\Re(\gamma_{kk})$, is the relative difference of \emph{with} interaction of the system-bath at each energy levels, instead of just difference of unperturbed eigenvalues, $\nu_1^D-\nu_2^D$. In our previous work~\cite{PKJ16+}, we confirmed that the crossings of self-energies of resonances, i.e., $\Re(\gamma_{jj})-\Re(\gamma_{kk})=0~\forall j,k$ take place in the region where the $d_B$ has extremal point with the same order of the quality factors of the modes. Therefore, the relative difference of the self-energies as the first part of the square root has zero value near $e=e_C$, and so the larger the relative difference of self-energies is getting, the farther it is from $e_C$. At the same time, $d_B$ at extremal point, i.e., the most similar decay channel  leads to the crossing of imaginary parts \big($\Im(\gamma_{11})=\Im(\gamma_{22})$\big). Now we focus on the relation between the wavefunctions' overlap and the $4\Re^2(\gamma')$ (i.e., the collective Lamb shift). The overlap is maximized near at the center of ARC ($e=e_{C}$) due to a mixing of wavefunctions by collective Lamb shift. Thus, this fact reveals that $4\Re^2(\gamma')$ (the off-diagonal component) is also maximized near at $e_{C}$. It indicates that these off-diagonal terms---an interaction \emph{via} the bath---become prominent in the regime of overlapping resonance~\cite{R09,R01,ER13}. From these results, we can conclude that the first term of the square root $[\Re(\gamma_{11})-\Re(\gamma_{22})]^2$ is minimized leading a crossing of imaginary part when the second term of the square root $4\Re^2(\gamma')$ is maximized, and there exist a trade-off relation between those two terms for the conservation of orthogonal relation in the total Hilbert space.

\section{Conclusion}
We have investigated a specific correlation between inner-part of and outer-part of a dielectric microcavity as a consequence in conservation law of orthogonal relations for the total (Hermitian) Hilbert space. This correlation is non-trivial because the non-Hermitian system with bi-orthogonal condition is always a subpart of the total Hilbert space. That is, the Hermitian total space can be decomposed into two orthogonal subspaces, one is a non-Hermitian system as an inner-part of the microcavity and the other is a resonance tail as an outer-part of the cavity lying on the bath. Furthermore, dealing with bath itself (extended to infinity) is also very complicated. This transition from non-Hermitian system to Hermitian system can explain that the inner product of wavefunctions in the non-Hermitian with different eigenvalues is non-zero in general, so that of resonance tails in bath must vary to remove this non-zero value for preserving the orthogonal relation in the total Hilbert space. The inner product of inner part of cavity is measured by overlap of wavefunctions whereas that of outer part of it is measured by Bhattacharra distance for below critical lines in Husimi distributions to overcome the difficulty of dealing with bath (infinity).

Finally, we applied this correlation to the Lamb shift and avoided crossing relating to the non-Hermitian Hamiltonian. As a result, we confirm that, on the contrary of the typical Hamiltonian model, in our study not only the off-diagonal-interaction terms but also the diagonal-interaction terms contributes to avoided crossings together under the given non-Hermitian Hamiltonian. The off-diagonal-interaction---the collective Lamb shift---which is related to the inner product of inner-part of the cavity is maximized at the center of ARC. However, the diagonal-interaction---the relative difference of self-energies---which is related to the outer-part of the microcavity is minimized (also leading to the crossing of imaginary part). The trade-off between these two terms determines the structure of the avoided crossing and it comes from the conservation of orthogonal relation in total space.

\section{Acknowledgments}
We are grateful to Kyungwon An and Ha-Rim Kim for valuable comments. This work was supported by a grant from Samsung Science and Technology Foundation under Project No. SSTF-BA1502-05. We thank Korea Institute for Advanced Study for providing computing resources (KIAS Center for Advanced Computation Linux Cluster) for this work. J.K. acknowledges financial support by the KIST Institutional Program (Project No. 2E26680-16-P025).
H.J. and K.J. also acknowledge financial supports by the National Research Foundation of Korea (NRF) through a grant funded by the Korea government (MSIP) (Grant No. 2010-0018295) and by the KIST Institutional Program (Project No. 2E26680-16-P025).
In addition, K.J. acknowledges financial support by the National Research Foundation of Korea (NRF) through a grant funded by the Ministry of Science and ICT (NRF-2017R1E1A1A03070510 and NRF-2017R1A5A1015626).

\section*{Appendix A. Derivation of the non-Hermitian Hamiltonian and Lamb shift} \label{NH-hamiltonian}
In this Appendix, we recapitulate the non-Hermitian quantum mechanics for an elliptic dielectric microcavity. First of all, let us consider a time-independent Schr\"{o}dinger equation with its total Hilbert space composed of two subsystems as follows:
\begin{equation} \label{eq:schrodinger}
\mathcal{H}_T\ket{\mu_T}_{SB}=\mu_T\ket{\mu_T}_{SB},
\end{equation}
where $\mathcal{H}_T$ is a total (Hermitian) Hamiltonian with (real energy) eigenvalues $\mu_T$, and $\ket{\mu_T}_{SB}$ represents the corresponding eigenvector of $\mu_T$ on a given (total) system $SB$. For convenience, each subspaces $S$ and $B$ denote a closed quantum system and a bath (or an environment), respectively.

The first subspace corresponds to a discrete state of closed quantum system $S$, and the second one is a continuous-scattering state of the bath $B$ such that projection operators, $p_S$ and $p_B$, satisfy $p_Sp_B=p_Bp_S=0$ and $p_S+p_B=\1_T$. Here, $p_S$ is a projection operator onto the closed quantum system whereas $p_B$ is a projection onto the bath. The operator $\1_T$ is an identity operator defined on the total space $SB$. With these projection operators, we can define useful matrices such as $H_S=p_S\mathcal{H}_Tp_S$, $H_B=p_B\mathcal{H}_Tp_B$, $\mathcal{V}:=V_{SB}=p_S\mathcal{H}_Tp_B$, and $\mathcal{V}^\dagger:=V_{BS}=p_B\mathcal{H}_Tp_S$.

The total Hamiltonian in equation~(\ref{eq:schrodinger}) can be represented by a matrix
\begin{equation} \label{eq:totalH}
\mathcal{H}_T=H_S+H_B+\mathcal{V}+\mathcal{V}^\dagger,
\end{equation}
where $H_S$ and $H_B$ denote the Hamiltonians of the closed quantum system and the bath with eigenvalues and eigenvectors such that $H_{S}\ket{\epsilon_{j}}_S=\epsilon_{j}\ket{\epsilon_{j}}_S$ and $H_{B}\ket{\beta}_B=\beta\ket{\beta}_B$, respectively. Here, $\mathcal{V}=V_{SB}$ and $\mathcal{V}^\dagger=V_{BS}$ are interaction Hamiltonians between the system and bath, respectively. The total eigenvector on $SB$ is also given by
\begin{equation}
\ket{\mu}_{SB}=\ket{\mu}_S+\ket{\mu}_B:=p_S\ket{\mu}_{SB}+p_B\ket{\mu}_{SB}.
\end{equation}
Notice that $p_S\ket{\mu}_{SB}=\ket{\mu}_S$ and $p_B\ket{\mu}_{SB}=\ket{\mu}_B$, and typically we have used usual addition `$+$' notation instead of the direct sum `$\oplus$'. (e.g. See the formalism as in Ref.~\cite{R09}.) By using these definitions, the Schr\"{o}dinger equation of equation~(\ref{eq:schrodinger}) can be rewritten~\cite{F58,R09} in the form of
\begin{align}
(H_B-\mu_T)\ket{\mu}_B&=-\mathcal{V}^\dagger\ket{\mu}_S\;\;\tn{and} \nonumber\\
(H_S-\mu_T)\ket{\mu}_S&=-\mathcal{V}\ket{\mu}_B.
\end{align}
Also note that the states restricted on $S$ and $B$ after the projections are given by~\cite{R09}
\begin{align}
\ket{\mu}_B&=\ket{\beta}_B+\mathcal{G}^{\triangleright}_B\mathcal{V}^\dagger\ket{\mu}_S\;\;\tn{and} \nonumber\\
\ket{\mu}_S&=\frac{\mathcal{V}}{\mu_T-\hat{H}_\tn{eff}}\ket{\beta}_B, \label{ev_A}
\end{align}
where $\ket{\beta}_B$---in general, it is called by homogeneous solution or plane wave---is the eigenvector of $H_B$, and $\mathcal{G}^{\triangleright}_B$ is an out going Green's function in the subspace $B$. The equation~(\ref{ev_A}) means that the eigenvector localized in subsystem $S$ can be obtained through the incoming vector $\ket{\beta}_B$ penetrating into the subsystem $S$ via coupling term $V_{SB}$ and propagating by effective Green's function $(\mu_T-\hat{H}_\tn{eff})^{-1}$.

Then, by using the total Hamiltonian in equation~(\ref{eq:totalH}), we can formally define the effective non-Hermitian Hamiltonian as
\begin{align}
\hat{H}_\tn{eff}&=H_S+\mathcal{V}\mathcal{G}_B^{\triangleright}\mathcal{V}^\dagger \\
&=H_S-\frac{1}{2}i\mathcal{V}\mathcal{V}^\dagger+P_\tn{v}\int\frac{\mathcal{V}\mathcal{V}^\dagger}{{\mu_T}-\tilde{\mu}_T}d\tilde{\mu}_T, \label{NHeff}
\end{align}
where $P_\tn{v}$ means the (Cauchy) principal value depending on each decay channels.

This non-Hermitian Hamiltonian has generally complex eigenvalues under the Feshbach projection-operator (FPO) formalism: (See Ref.~\cite{PKJ16} and the references therein.)
\begin{equation} \label{eq:Heff}
\hat{H}_\tn{eff}\ket{\psi_{j}}_\tn{eff}=\nu_{j}\ket{\psi_{j}}_\tn{eff}
\end{equation}
and its complex energies (eigenvalues) of $\hat{H}_\tn{eff}$ are given by (for each $j$)
\begin{equation} \label{eq:evHeff}
\nu_j=\mu_j-\frac{i}{2}\omega_j.
\end{equation}
Here, $\nu_{j}$ is an eigenvalue relating to $\mu_j$ and $-\tfrac{\omega_j}{2}$, which we call just energy (real part) and decay-width (imaginary part) of $j$-th eigenvector~\cite{R09,CK09,D00,WKH08}, respectively.

Now we consider the Lamb shift. The Lamb shift, which is a small energy difference between a closed and an open system due to the system-bath interaction, can be also obtained by the effective non-Hermitian Hamiltonian in equation~(\ref{NHeff})~\cite{PKJ16,PKJ16+,R13}. That is, it is the difference between the eigenvalues of $H_S$ and the real part of the eigenvalues of $\hat{H}_\tn{eff}$:
\begin{equation} \label{eq:defLamb}
L_{j}:=\mu_{j}-\epsilon_{j}.
\end{equation}
In general, while $H_S$ itself has real eigenvalues, the effective non-Hermitian Hamiltonian has complex energy eigenvalues. Thus we consider a complex matrix as followings:
\begin{equation}
\Gamma:=\hat{H}_\tn{eff}-H_S=\begin{pmatrix}
\gamma_{11} & \gamma_{12} \\
\gamma_{21} & \gamma_{22}
\end{pmatrix}.
\end{equation}
If we consider the case of a two-dimensional vector space, the non-Hermitian Hamiltonian with respect to a given eigenbasis of $H_{S}$, i.e., $\{\ket{\epsilon_j}_S\}_{j=1}^2$ can be represented by $2\times2$ toy-Hamiltonian-model as follows ($\hat{H}_\tn{eff}=H_S+\Gamma$):
\begin{align}
\hat{H}_\tn{eff}=
\begin{pmatrix}
\epsilon_1+\gamma_{11} & \gamma_{12} \\
\gamma_{21} & \epsilon_2+\gamma_{22}
\end{pmatrix}.
\end{align}
For more additional works, above equation can be reordered as (Note that $D$ and $V$ below indicate diagonal and off-diagonal matrix of the effective Hamiltonian, respectively.)
\begin{equation}
\hat{H}_\tn{eff}=(\hat{H}_{\tn{eff}})_D+(\hat{H}_{\tn{eff}})_{V},
\end{equation}
where each components have following forms (actually this division of $\hat{H}_\tn{eff}$ conforms our main suggestion)
\begin{align}
(\hat{H}_{\tn{eff}})_D
=\begin{pmatrix}
\epsilon_{1}+\gamma_{11} & 0 \\
0            & \epsilon_{2}+\gamma_{22}
\end{pmatrix}
\end{align}
and
\begin{align}
(\hat{H}_{\tn{eff}})_{V}
&=\begin{pmatrix}
0 & \gamma_{12} \\
\gamma_{21} & 0
\end{pmatrix},
\end{align}
where $\epsilon_j+\gamma_{jj}:=\mu_j-\frac{i}{2}\omega_j~\forall j\in\{1,2\}$, thus we have the real value $\Re(\gamma_{jj})=\mu_j-\epsilon_j$ and the imaginary value $\Im(\gamma_{jj})=-\frac{1}{2}\omega_j$ for any complex number $\gamma_{jj}$.

In Refs.~\cite{SS10,R13}, the term `self-energy' due to the diagonal components $\Re(\gamma_{jj})$ was known as `Lamb shift' in atomic physics. On the other hand, the off-diagonal terms $\gamma_{jk}\;(\forall j,k=\{1,2\})$ were known as `collective Lamb shift'. That is, the self-energy is an energy shift due to the individual energy-level interactions with the bath independently, while due to the energy-level interactions via the bath each other, is called the collective Lamb shift by Rotter~\cite{R13}.

\section*{Appendix B. Bi-orthogonality condition} \label{Bi-orthogonality}
For convenience, let us abbreviate $\hat{H}_\tn{eff}$ as
\begin{equation}
\hat{H}_\tn{eff}=H-i\Omega,
\end{equation}
where $H^\dagger=H$ and $\Omega^\dagger=\Omega$ satisfying non-Hermitian Hamiltonian with eigenvectors and eigenvalues in the form of
\begin{equation}
\hat{H}_{\tn{eff}}\ket{\psi_{j}}=\nu_{j}\ket{\psi_{j}}\:\:\tn{and}\:\:
\bra{\psi_{j}}\hat{H}^{\dag}_{\tn{eff}}=\nu_{j}^{*}\bra{\psi_{j}}.
\end{equation}
We also assume that the eigenvalues of $\hat{H}_\tn{eff}$ are not degenerate. In addition to the eigenvector of $\hat{H}_\tn{eff}$, it will be convenient to invoke an eigenvector $\ket{\phi_j}$ of the Hermitian adjoint matrix $\hat{H}_\tn{eff}^{\dag}$ in the form of
\begin{equation}
\hat{H}_\tn{eff}^{\dag}\ket{\phi_{j}}=\kappa_{j}\ket{\phi_{j}}\:\:\tn{and}\:\:
\bra{\phi_{j}}\hat{H}_{\tn{eff}}=\kappa_{j}^{*}\bra{\phi_{j}}.
\end{equation}

The reason for invoking the additional states $\ket{\phi_j}$ is since the eigenvector $\ket{\psi_{j}}$ of $\hat{H}_{\tn{eff}}$ are not orthogonal in general.
This result can be conformed by noting the facts:
\begin{align}
2i\Omega=\hat{H}_{\tn{eff}}^{\dag}-\hat{H}_{\tn{eff}}\:\:\tn{and}\:\:
2H=\hat{H}_{\tn{eff}}^{\dag}+\hat{H}_{\tn{eff}}.
\end{align}
Therefore, the identity operators are given by
\begin{align}
\1=\frac{2i\Omega}{\hat{H}_\tn{eff}^{\dag}-\hat{H}_\tn{eff}}\:\:\tn{and}\:\:
\1=\frac{2H}{\hat{H}_\tn{eff}^{\dag}+\hat{H}_\tn{eff}},
\end{align}
where we note that, in our case, the effective Hamiltonian has always full-rank.
By using this relations, we will get the following results:
\begin{align}
\inn{\psi_{j}}{\psi_{k}}
=2i\frac{\bra{\psi_{j}}{\Omega}\ket{\psi_{k}}}{\nu_{j}^{*}-\nu_{k}}
=2\frac{\bra{\psi_{j}}{H}\ket{\psi_{k}}}{\nu_{j}^{*}+\nu_{k}}.
\end{align}
Due to the fact that the vector $\ket{\psi_{j}}$ is an eigenvector of $\hat{H}_{\tn{eff}}$ not of $\Omega$ and $H$, the numerators $\bra{\psi_{j}}{\Omega}\ket{\psi_{k}}$ and $\bra{\psi_{j}}{H}\ket{\psi_{k}}$ are non-zero in general.
As a result, the inner product $\inn{\psi_{j}}{\psi_{k}}$ are non-zero~\cite{B14}.
So an analogous result can be found as
\begin{align}
\inn{\phi_{j}}{\phi_{k}}
=2i\frac{\bra{\phi_{j}}{\Omega}\ket{\phi_{k}}}{\nu_{j}-\nu_{k}^{*}}
=2\frac{\bra{\phi_{j}}{H}\ket{\phi_{k}}}{\nu_{j}+\nu_{k}^{*}}.
\end{align}

The eigenvectors with different eigenvalues belonging to non-Hermitian matrices are not orthogonal. Thus, it is natural and desired to establish an orthogonal conditions again. To do so, let us introduce `right' and `left' eigenvectors of the non-Hermitian Hamiltonian, which are different from each other in general. First of all, if we consider equation~(\ref{eq:Heff}) again
\begin{align}
\hat{H}_{\tn{eff}}\ket{\psi^{R}_{j}}=\nu_{j}\ket{\psi^{R}_{j}},
\end{align}
then, by multiplying to the left-hand side with left eigenvectors $\bra{\psi^{L}_{k}}$, one obtains that
\begin{align}
\bra{\psi^{L}_{k}}\hat{H}_{\tn{eff}}\ket{\psi^{R}_{j}}&=\nu_{j}\inn{\psi^{L}_{k}}{\psi^{R}_{j}}
=\nu_{j}\delta_{jk},
\end{align}
where the right and left eigenvectors are assumed to be orthonormal. Thus we can easily check that
\begin{align}
\bra{\psi^{L}_{j}}\hat{H}_{\tn{eff}}=\nu_{j}\bra{\psi^{L}_{j}},
\end{align}
which is corresponding to the Schr\"{o}dinger equation for the left bra-vectors. If we apply a dagger operation to this vector, we get
\begin{align}
\hat{H}^{\dag}_{\tn{eff}}\ket{\psi^{L}_{j}}=\nu_{j}^{*}\ket{\psi^{L}_{j}}.
\end{align}

In the case of Hermitan Hamiltonian operator, that is, $\hat{H}^{\dag}_{\tn{eff}}=\hat{H}_{\tn{eff}}$, it directly lead to $\nu_{j}$ is real and $\ket{\psi^{L}_{j}}=\ket{\psi^{R}_{j}}$. However, when $\hat{H}_{\tn{eff}}$ is not the case of Hermitian Hamiltonian, i.e., $\hat{H}^{\dag}_{\tn{eff}}\neq \hat{H}_{\tn{eff}}$, furthermore, if $\hat{H}_{\tn{eff}}$ is symmetric,
\begin{align}
\hat{H}^{\dag}_{\tn{eff}}=\hat{H}^{*}_{\tn{eff}},
\end{align}
then, this gives birth to the relation between the left and right eigenvectors
\begin{align}
\ket{\psi^{L}_{j}}=\ket{\psi^{R}_{j}}^{*}.
\end{align}
This is a typical physical system because time reversal condition is usually satisfied~\cite{L09}. In fact, by comparing the previous results, we can notice that
\begin{align}
\ket{\psi^{R}_{j}}\rightarrow\ket{\psi_{j}}\:\:\tn{and}\:\:
\ket{\psi^{L}_{k}}\rightarrow\ket{\phi_{k}}.
\end{align}
This bi-orthogonal relation is dominant as approaching to the exceptional point, whereas nearly orthogonal ($\hat{H}_\tn{eff}\approx H_S$) away from the exceptional point in non-Hermitian Hamiltonian~\cite{R09}.

\begin{figure*}
\centering
\includegraphics[width=17cm]{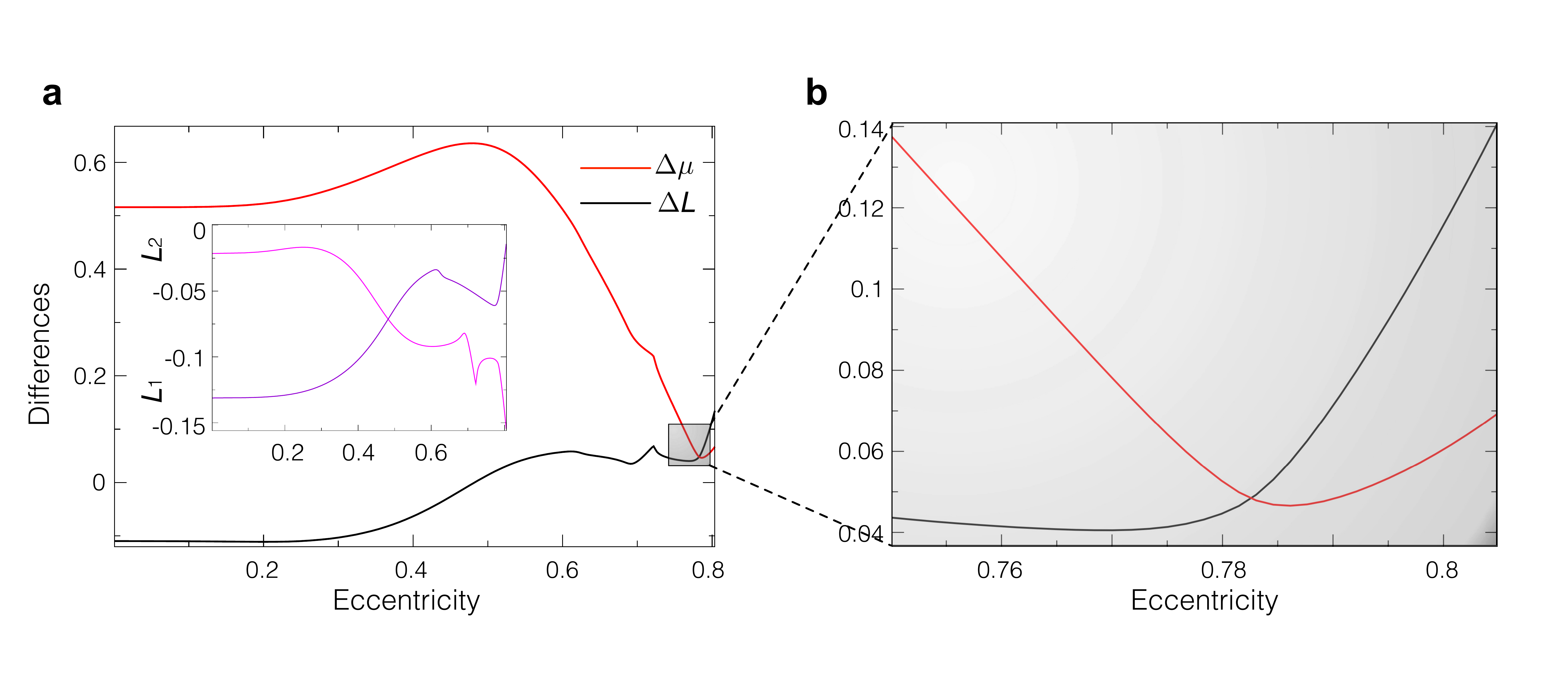}
\caption {{\bf The relation between relative difference of energies and that of Lamb shifts}. (\textbf{a}) There are Lamb shift, $L$, inside of \textbf{a} as an inset obtained from relative difference between adjacent thick grey and thick brown solid curves in Fig.~\ref{Figure-1}. The black curve is a difference between two Lamb shift of inset and the red curve is a that of between two thick brown curves in Fig.~\ref{Figure-1}\textbf{b}. (\textbf{b}) The enlarged region of shadow box in Fig.~\ref{Figure-5}\textbf{a}. The red and black solid curves meet at $e\approx0.782$ as predicted in Eq.~(\ref{eq:eequivL}) below.}
\label{Figure-5}
\end{figure*}

\section*{Appendix C. Relation between Lamb shift and ARC} \label{LambARC}

To find out clues of relation between Lamb shift and ARC, let us consider relative difference between a pair of Lamb shift $(L_1,L_2)$ and that of the real-valued energies in eigenvalue trajectories $(\mu_1,\mu_2)$ for microcavity depending on ${e}$ simultaneously. To do that, we define $\mu_{1,2}$ as follows:
\begin{align}
\mu_{1,2}(e)&=\epsilon_{1,2}+L_{1,2}.
\end{align}

Then $\mu_{1,2}$ corresponds to thick brown solid curves in Fig.~\ref{Figure-1}\textbf{a} and $\epsilon_{1,2}$ do to thick grey solid curves in Fig.~\ref{Figure-1}\textbf{a}, respectively. Their Lamb shift $L_j$ for each $j$ are shown by the inset of Fig.~\ref{Figure-5}\textbf{a}. The violet solid curve is a $L_1$ which is a Lamb shift of upper adjacent thick grey and thick brown solid curves in Fig.~\ref{Figure-1}\textbf{a} while magenta solid one is a $L_2$ which is a that of lower adjacent thick grey and thick brown solid curves, respectively. Then we can define the relative difference of Lamb shift depending on the eccentricity $e$ as
\begin{align}
\Delta{L}(e)=L_1-L_2.
\end{align}
This one is depicted as a black solid curve in Fig.~\ref{Figure-5}\textbf{a}. Similarly,
\begin{align}
\Delta\mu(e)
&=\epsilon_1-\epsilon_2+\Delta L.
\end{align}
This one is a red solid curve depicted in Fig.~\ref{Figure-5}\textbf{a}. 

At the center of ARC ($e=e_C$), we get
\begin{align}
\Delta\mu(e_C)&=\epsilon_2(e_C)-\epsilon_1(e_C)+\Delta L(e_C).
\end{align}
It means that the real-valued energies difference for open quantum system consists of two part, i.e., first one is a relative difference of the real-valued energies of the billiard and second one is that of the Lamb shift. The center of ARC exits around crossing point in general, but in order to get deeper understanding clearly, let us consider the circumstance when two points are almost coincident ($e_{X}\approx e_{C}$); thus, $\epsilon_1\approx\epsilon_2$. Then the difference of real-valued energies $\Delta\mu(e_{C})$ for open system is exact difference of Lamb shift: At the eccentricity $e=e_C$,
\begin{align} \label{eq:eequivL}
\Delta\mu& \approx \Delta L.
\end{align}
These are clearly shown in Fig.~\ref{Figure-5}\textbf{b}. The crossing point ($e_{X}$) and the center of ARC ($e_{C}$) are same in $e\approx0.782$ so that $\Delta\mu(e_{C})$ is equal to $\Delta L(e_{C})$ with $\approx0.05$.


\end{document}